\documentclass[twocolumn]{aastex631}
\usepackage{amsmath}
\usepackage{ulem}
\usepackage{soul}
\usepackage{todonotes}
\usepackage{graphicx}

\newcommand{\yt}[1]{\textcolor{black}{#1}}

\begin{document}

\title [LSST CV discovery]{Discovering Cataclysmic Variables from the Rubin Observatory LSST}

\author{D. A. H. Buckley $\dagger$}
\affiliation{South African Astronomical Observatory, PO Box 9, Observatory, 7935, Cape Town, South Africa}
\affiliation{Department of Physics, University of the Free State, PO Box 339, Bloemfontein 9300, South Africa}
\affiliation{Department of Astronomy, University of Cape Town, Private Bag X3, Rondebosch 7701, South Africa}
\email{dibnob@saao.ac.za}
\correspondingauthor{D. A. H. Buckley \\ $\dagger$First and second author contributed equally.}

\author{Y. Tampo $\dagger$}
\affiliation{South African Astronomical Observatory, PO Box 9, Observatory, 7935, Cape Town, South Africa}
\affiliation{Department of Astronomy, University of Cape Town, Private Bag X3, Rondebosch 7701, South Africa}

\author{P. Szkody}
\affiliation{Department of Astronomy, University of Washington, Seattle, WA 98195-1700, USA}

\author{M. Motsoaledi}
\affiliation{Department of Astronomy, University of Cape Town, Private Bag X3, Rondebosch 7701, South Africa}

\author{S. Scaringi}
\affiliation{Centre for Extragalactic Astronomy, Department of Physics, Durham University, South Road, Durham, DH1 3LE, UK}
\affiliation{INAF-Osservatorio Astronomico di Cap, Salita Moiariello 16, I-80131 Naples, Italy}

\author{M. Lochner}
\affiliation{Department of Physics and Astronomy, University of the Western Cape, Bellville, Cape Town, 7535, South Africa}

\author{N. Rawat}
\affiliation{South African Astronomical Observatory, PO Box 9, Observatory, 7935, Cape Town, South Africa}

\author{J. P. Marais}
\affiliation{Department of Physics, University of the Free State, PO Box 339, Bloemfontein 9300, South Africa}

\author{B. van Soelen}
\affiliation{Department of Physics, University of the Free State, PO Box 339, Bloemfontein 9300, South Africa}

\author{S. Macfarlane}
\affiliation{South African Astronomical Observatory, PO Box 9, Observatory, 7935, Cape Town, South Africa}
\affiliation{Department of Astronomy, University of Cape Town, Private Bag X3, Rondebosch 7701, South Africa}

\author{A. van Dyk}
\affiliation{South African Astronomical Observatory, PO Box 9, Observatory, 7935, Cape Town, South Africa}
\affiliation{Department of Astronomy, University of Cape Town, Private Bag X3, Rondebosch 7701, South Africa}



\begin{abstract}

The Vera C. Rubin Observatory's Legacy Survey of Space and Time (LSST) will provide a windfall of new transients and variable sources.
Here we have performed mock observation simulations to understand LSST's expected detection rates for cataclysmic variables (CVs) with known large amplitude variations.
{Under the thin-disk approximation for the distribution of CVs in our Galaxy, }we found that {only 20\%} of WZ Sge-type dwarf novae systems, representing the most energetic disk-driven outbursts in CVs, will be detected during outbursts by the LSST. Given their large amplitude (7--9 mag), {only those} brighter than $17.5$ mag at outburst maximum are expected to have an $r$-band quiescence counterpart in individual scans. Thanks to the {planned} cadence of the LSST {towards the Galactic center}, {$\approx$70\%} of the simulated outbursts will be detected twice or more on the discovery night, and two-thirds will be observed in different bands.
CVs of the Polar class, which display luminosity changes up to 4 mag, can be unbiasedly recovered to 22.5 mag with more than 100 detections over 10 years of the LSST operation. Finally, we attempt to characterize the detection rate of micronovae bursts, and find that about {2.6\%} of the simulated sample will be observed as a $\geq 0.4$ mag-amplitude and $\leq$ 1-d duration spike in the long-term light curve.
Overall, our results consolidate LSST's capability to studying time-domain phenomena in CVs, and inform on how to plan and organize follow-up observation strategies on transients discovered by LSST.

\end{abstract}



\section{Introduction}

Cataclysmic variables (CVs) are close binary stars consisting of an accreting white dwarf primary and a lower mass late type (typically G-L spectral type main sequence) secondary star, which fills its Roche Lobe and transfers mass to the primary. In the absence of a strong magnetic field, conservation of angular momentum and viscosity forces the material into an accretion disk around the white dwarf. The viscous forces cause the accreted material to slowly spiral in toward the white dwarf and settle onto its surface \citep[see, e.g.][]{Warner1995}.

By their nature, CVs vary in their brightness over a range of timescales and amplitudes. Powered by accretion, CVs show stochastic variability, referred to as ``flickering", as result of the turbulence in the accretion flow onto the accreting white dwarf, whether via an accretion disk or directly, in the case of magnetic systems. Power density spectra usually reveal a characteristic power law shape, consistent with that expected from Kolmogorov turbulence. Flickering is typically seen on timescales of minutes down to sub-seconds.

Larger amplitude variations in CVs, of which their detection in the LSST is the subject of this paper, come about through changes in accretion rate, which can occur on timescales of hours, days or weeks.  The various sub-classes of CVs considered here show brightness changes of between factors of $\sim 10^1-10^4$.
Dwarf novae (DNe) represent the largest sub-group of cataclysmic variables \citep[e.g.; $\approx$60\% of 150-pc samples in][]{pal20GaiaCVdensity} and are expected to be one of the major sources of transients in the LSST survey \yt{ \citep{rid14surveys}. They represent the largest predicted classes of Galactic transient sources, judging from the results of the Catalina Real Time Survey \citep[CRTS][]{Drake2009}.} 
The most comprehensive model for DN outbursts comes from the disk instability model \citep[DIM, e.g.][]{Osaki1974, hos79DImodel, Meyer1981, Cannizzo1982, Smak1984}. In this model, the accretion disk develops thermal instability when the disk reaches the critical surface density and hydrogen gas in the disk ionizes. A propagation front of the heating wave develops and moves radially inward and outward from the instability, increasing the accretion rate in the disk and onto the white dwarf. A recent discussion of the DIM can be found in \cite{kim20thesis, Hameury2020}.

Modern all-sky surveys are discovering new CVs and DNe at unprecedented rates, such as the Catalina Real Time Survey \citep[CRTS; ][]{Drake2009} and the Zwicky Transient Facility \citep[ZTF; see e.g.][]{Graham2019, Szkody2020, Szkody2021}. Although there are many efforts to conduct prompt follow-up observations, especially for bright systems \citep[e.g.][]{kat09pdot, tam21dnfollowup}, a certain fraction of DNe are discovered in the data after the initial outburst, which prevents detailed followup studies of the outbursts at suitable cadences.

The upcoming Vera C. Rubin Legacy Survey of Space and Time (LSST), with its large collecting area, wide field-of-view, high cadence and multi-filter observation strategy will allow an unprecedented opportunity to discover and study new and exciting transient events. It will provide an opportunity to quickly detect new transients still in their early stages and gain a deeper understanding of the evolution of transient classes in time. 
In an earlier investigation, we produced an LSST "Cadence Note"
\citep[][]{cadence_note} with {the LSST Operations Simulator \citep[OpSim;][]{Delgado2014, Delgado2016, Reuter2016}} baseline v1.7, where we looked at simulations of dwarf nova light curves informed by results from CRTS \citep{Drake2009}, an unfiltered transient detection survey. There we considered the parameters characterizing these light curves, including the amplitude of the outburst and the outburst length, which were in the range $2 < \Delta\mathrm{m}_g < 8$ and $7 < t < 20$ days respectively. 
In this paper, we investigate more types of CVs with various characteristic timescales using realistic light curves {and on-sky distributions}, and with the latest version of OpSim, namely baseline v4.3.1.

In section \ref{sec:simulation} we present simulations of light curves of three distinct classes of cataclysmic variables, namely: 1.) \yt{large-amplitude DN outbursts} of the WZ Sge sub-type, 2.) magnetic cataclysmic variables of the polar sub-type, which switch between different brightness states, and 3.) fast micronova eruptions. These three groups represent a range of outburst amplitudes and timescales. We use real example light curves of a number of these systems to create simulated LSST light curves based on the simulated Opsim cadence. We consider two on-sky distributions to study the discovery spaces of the Galactic disk and other populations of CVs.
We summarize the results of our simulations in section \ref{sec:result} and present the discovery space of the given samples. Finally, in section \ref{sec:discussion}, we discuss the capability of the LSST both for CVs and other transients with similar timescales{, and give a summary of the paper in section \ref{sec:summary}}.

\section{LSST simulation of cataclysmic variables}
\label{sec:simulation}

\subsection{LSST Opsim Simulator}
\label{sec21:opsim}

\yt{The} Opsim environment is a python package that is used to simulate LSST observations over the planned 10 year LSST of the southern sky. The OpSim uses the science program requirements, telescope design and modeled environmental conditions to produce an accurate simulation of observation schedules and conditions over the 10 year period of the survey. The OpSim database is a vital component for any studies regarding the planned survey. We used information regarding the planned pointing schedule to simulate a realistic cadence for the light curves.  Other uses for the OpSim database include planned coverage, visibility and expected source detection and measurement studies\footnote{\url{https://www.lsst.org/scientists/simulations/opsim}}.

We used the latest version of the survey baseline v4.3.1\footnote{\url{https://survey-strategy.lsst.io/baseline/index.html}}, which includes the Main Survey (the Wide Fast Deep; WFD), Mini Survey, Micro Survey, Deep Drilling Fields (DDF), and simulated Target of Opportunity (ToOs) observations. The typical 5-$\sigma$ limiting magnitudes with a single visit (2 $\times$ 15 s) are $g\approx24.3$ and $r\approx24.0$. The WFD observations, which compose more than 80\% of the observation time, are characterized by around 800 visits per pointing over 10 years {including both low-dust extinction areas and higher stellar density areas towards the Galactic center} (see the left-top panel of figure \ref{fig:yt-skydist-TD}). The entire visible sky will be uniformly observed in Year 1, while in later years a rolling cadence, alternating high and low activity areas on the sky each year, is currently planned {for the low-dust extinction regions}. In an active rolling cadence season, the survey cadence will be 2--4 days, while in inactive seasons the cadence will be lower. Typically there will be two visits at a given field, in different filters, on a given night, with repeat visits typically involving other filters.  
\yt{This will allow the detection of some fast rising/declining transients, with timescales of $\lessapprox$ 0.3 d, although for longer timescales ($\sim0.5-2$ d), their rising/declining timescales will not be well constrained due to $\geq2$-d cadences.} The DDFs cover only a small fraction of the sky but will be visited much more often; $\approx$23000 visits in 10 years and 10--20 visits per filter within a single night.

\subsection{Input light curves}
\label{sec22:inputlc}

\begin{table*}[]
    \caption{Summary of input light curves}
    \centering
    \begin{tabular}{cccccccc}
    \hline
    Name & Type  & $M_{r~\rm max}$ & $M_{r~\rm min}$ & Color$_{\rm max}$ & Color$_{\rm min}$ & Light curve reference \\
     & &  [mag] & [mag] & & & \\
    \hline
    \hline
    WZ Sge & UGWZ  & 3.86 & 11.95 & 12000 K BB& Pan-STARRS1 & \citet{kat09pdot} \\
    V455 And & UGWZ  & 4.25 & 11.47 & 12000 K BB & Pan-STARRS1  & \citet{kat09pdot} \\
    GW Lib & UGWZ  & 3.08 & 11.35 & 12000 K BB & Pan-STARRS1  & \citet{kat09pdot} \\
    BM CrB & Polar & 8.20 & 12.36 & 12000 K BB & 12000 K BB &  \citet{Drake2009}\\
    AR UMa & Polar & 9.50 & 11.50 & 12000 K BB & 12000 K BB &  \citet{Drake2009}\\
    V808 Aur & Polar & 7.80 & 11.68 & 12000 K BB & 12000 K BB &  \citet{Drake2009}\\
    ASASSN-19bh & micronova & 1.75 & 5.10 & 12000 K BB & ATLAS-REFCAT2 & \citet{sca22micronova} \\
    TV Col $\#$1 & micronova & 3.52 & 5.49 & 12000 K BB & ATLAS-REFCAT2 & \citet{sca22micronova} \\
    TV Col $\#$2 & micronova & 3.21 & 5.49 & 12000 K BB & ATLAS-REFCAT2 & \citet{sca22micronova} \\
    \hline
    \end{tabular}
    \label{tab:inputlcs}
\end{table*}

\begin{figure*}[t]
    \centering
    \includegraphics[width=\textwidth]{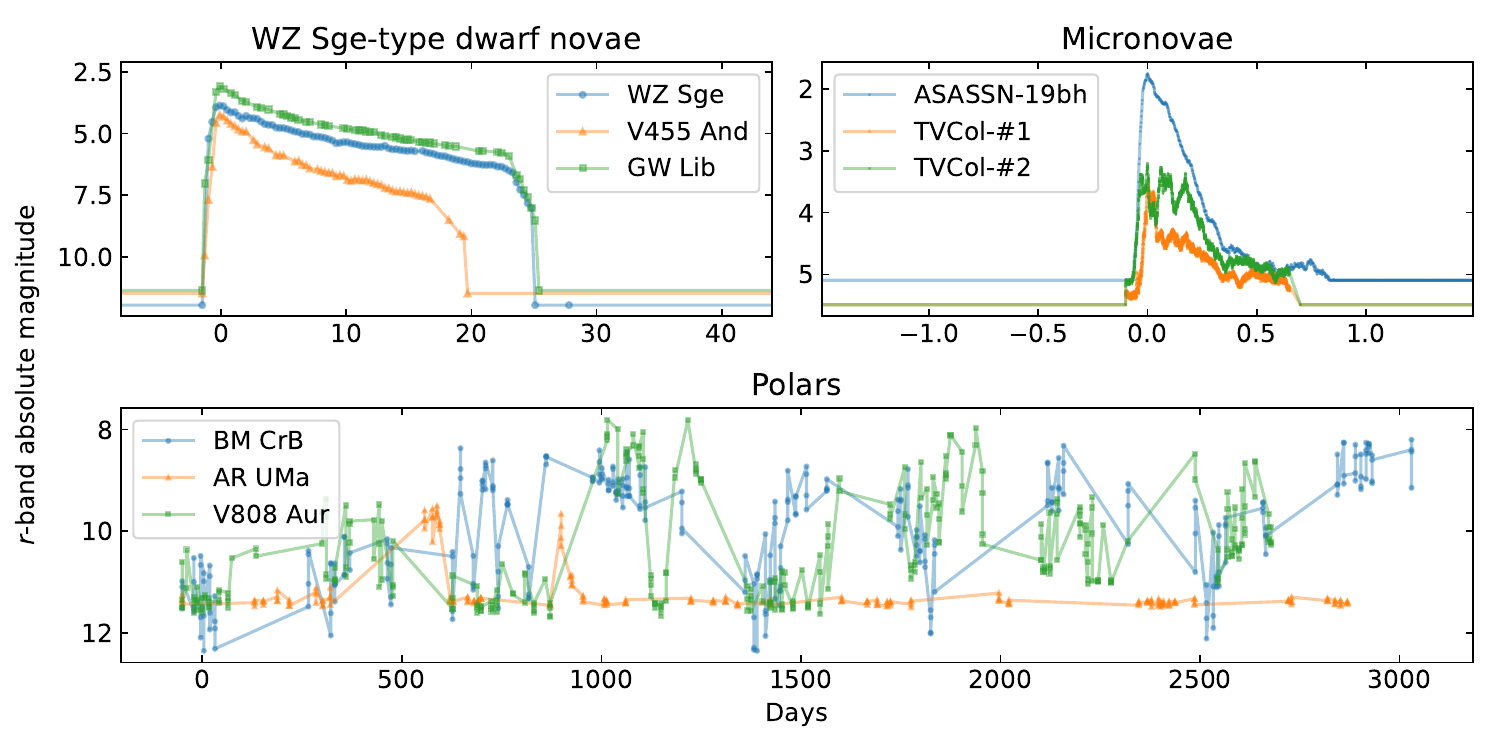}
    \caption{$r$-band input light curves of WZ Sge-type DNe (top left), micronovae (top right), and polars (bottom), in the absolute magnitude unit.}
    \label{fig:yt-tmpllc}
\end{figure*}

As mentioned in the Introduction, we simulate WZ Sge-type DNe, multi-state polars, and micronova bursts to cover a range of timescales in CVs. Figure \ref{fig:yt-tmpllc} presents the $r$-band light curves of our samples in the absolute magnitude unit. The details of compiling these light curves for our simulation are described below on each class. Once the template light curves are obtained, they are scaled to absolute magnitude using the Gaia EDR3 distances \citep{Bai21GaiaEDR3distance}. Finally, the simulations are performed by interpolating the input light curve linearly on a magnitude scale in each band.

\subsubsection{WZ Sge-type dwarf novae}
\label{sec221:wzsge}

WZ Sge-type DNe show the lowest mass-transfer rate, largest outburst amplitude (up to 9 mag), and longest outburst cycles (typically decades) among DNe \citep[see a review in][]{kat15wzsge}. Thus compared to other DNe with short outburst cycles, only one outburst from each system is expected even during the entire 10-year LSST operation. At the same time, WZ Sge-type DNe are an important population in CVs. The population studies suggest that these low-mass transfer rate systems should dominate the CV population \citep[40--70\%; e.g.][]{kol93evolution} while the observed fraction is less than a quarter in the volume-limited samples \citep{pal20GaiaCVdensity}. The wide and deep coverage of the LSST can increase its sample by a large factor to comprise more deep insights on this problem.
\yt{
It is worth noting that the brightness of DNe in quiescence and outburst cycle strongly differs between subtypes, although their peak brightness is more similar to each other \citep[e.g.][]{abr22gaiaCV}.  Hence, our cases of WZ Sge stars, which are characterized by the faintest in quiescence and have the longest outburst cycles, should be considered as the lowest detection efficiency among various subtypes in DNe. }


In order to properly treat the fast-evolving light curve of WZ Sge-type DNe in outburst, we used the well-sampled light curves of WZ Sge, GW Lib, and V455 And in outburst observed through the VSNET collaboration \citep{kat09pdot}. {These three systems are covered from very early in their outburst rise.} 
\yt{The original data are observed with a clear band and calibrated in the V-band magnitude (CV magnitude). We use this} light curve as the $g$-band light curve in our simulation. We binned the data by 0.3 d to remove the orbital variations. \yt{For other bands, we determined their color difference from the $g$ band applying} the blackbody color of 12000 K, which is the typical color observed in WZ Sge-type DNe in outburst \citep[e.g.][]{mat09v455and}.  
For quiescence before and after the outburst, we applied the magnitudes from the Pan-STARRS1 catalog \citep{cha16ps1} except for the $u$ band, which is unavailable. We assumed $u$ = $g$ in AB mag.
Although WZ Sge stars show a post-outburst decline lasting about a year and a series of rebrightenings, we assume that the system returns to quiescence level after the rapid decline from the outburst to simplify the input light curve.

\subsubsection{Polars}
\label{sec222:polar}

The polar class of magnetic cataclysmic variable involves direct accretion of magnetically funneled material, rather than passing through an accretion disk. As such, changes in mass loss rate through the L$_1$ point of the companion's Roche lobe leads to directly correlated changes in the accretion luminosity. 
Polars show variations in their optical brightness, typically between "high'' and "low" states, separated by 2--4 magnitudes. Results from the extensive CRTS observations \citep{Drake2009}  have established that the timescales for the duration of these states can be from months to years \citep{khinethesis}. 


We used the CRTS light curve of polars BM CrB, AR UMa, and V808 Aur. 
BM CrB and V808 Aur are relatively active systems with multiple state-changes over the CRTS observations. On the other hand, AR UMa mostly stays in its low state and shows only two short-lived high states. Moreover, V808 Aur is shown to have 3 states; high, middle and low \citep{khinethesis}. Compared to our WZ Sge star sample, the sampling cadence of CRTS is much sparser (typically $\geq$ 10 d and includes seasonal gaps). Nevertheless, as our main purpose is to demonstrate the capability of LSST to recover the different states in polars, we used the interpolated light curve as input light curves. Since CRTS only observes with a clear-band filter, we treated their raw data as $g$-band light curve and assumed the blackbody color of 12000 K in all states \yt{to create the input light curves in other bands}.

\subsubsection{Micronovae}
\label{sec223:micronovae}

Micronovae are a newly identified phenomenon in \yt{some} intermediate polars (IPs) of magnetic CVs \citep{sca22micronova}. These bursts are characterized by a few-magnitude brightening lasting only for one day or less. The recurrence time is estimated $\sim 1$ year \citep{sca22micronova}. The suggested interpretation is localized thermonuclear bursts on the WD. 
We used the light curves of ASASSN-19bh and TV Col in TESS, scaled to the $g$-band magnitude of ASAS-SN in the same manner as \citet{sca22micronova}. We assumed a 12000-K blackbody color during the burst for simplicity, while we adopted the colors in the ATLAS all-sky stellar reference catalog \citep[ATLAS-REFCAT2; ][]{Atlasrefcat2} in quiescence. We assumed $u=g$ and $y=z$ in AB magnitude scale as $u$ and $y$ bands are not available in ATLAS-REFCAT2. 
In \yt{the} case of ASASSN-19bh, due to its intrinsically red color in quiescence, the effective duration of an eruption is shorter in redder filters.

\subsection{On-sky distribution}
\label{sec23:skydist}

We simulated 20000 light curves in each class, 
\yt{using} the above-mentioned sample light curves.
This is mainly limited by the computational power; the number of CVs in our galaxy is expected to be an order of millions based on the space density of local CVs \citep{lsstsciencebookv2}.
For WZ Sge stars and micronovae, the epoch of outburst maximum was randomly chosen during the first two years of the LSST operating period ({\texttt{rubin\_scheduler.utils.SURVEY\_START\_MJD}; }MJD 60980.0 {(= 2025 November 1st UT)} -- 61710.0), including in solar conjunction, to verify the performance of the LSST in its initial years.
For polars, in order to avoid the simulator observing only the seasonal gaps in the input light curves, we manually shifted the input light curve so that the seasonal gaps in CRTS would match with those in the LSST at the given RA of each sample. 
\yt{We consider two models of the on-sky distribution of our coordinate inputs: the thin-disk and uniform distributions. We note that our uniform distribution does not represent any realistic distribution of CVs, however, it is still useful to constrain the general performance of LSST on CV studies beyond and above the Galactic plane (e.g., in halos, globular clusters, and even nearby galaxies) as seen in section \ref{sec35:uniform_model}.} 
The 3-D extinction map \texttt{DustMap3D} implemented in the LSST OpSim was used to calculate the Galactic extinction at the given RA, Dec, and distance.

\subsubsection{{Thin disk distribution}}
\label{sec232:thindisk}

\yt{The distribution of CVs in our galaxy has been approximated as the thin disk population \citep[e.g.][]{pre07CV_PGsample}, as the CV population should loosely follow the Galactic star formation history \citep[Galactic classical novae for example;][]{ozd18galacticnova}.}
We follow the approach of \citet{pre07CV_PGsample}; the thin disk is assumed to be axisymmetric with no spiral structure, no thick disk, no halo, and no Galactic bulge. We produced the celestial position randomly at the radial and vertical distributions following the equations \ref{eq:Hz} and \ref{eq:Hr}, where $(r, z, \phi)$ is a cylindrical coordinate system centered at the Galactic center. 

\begin{align}
    \rho (z) \propto \exp{- |z| / H_z} \label{eq:Hz}\\
    \rho (r) \propto \exp{- r / H_r}  \label{eq:Hr}
\end{align}

{The Earth is placed at $(r, z, \phi)$ = (7620 pc, 0, 0), and we applied radial scale height $H_r = 3000$ pc following \citet{pre07CV_PGsample}. The vertical scale height $H_z$ is considered to vary depending on the CV population, and \citet{pre07CV_PGsample} introduced 120, 260, and 450 pc for long-$P_{\rm orb}$ (i.e.; above period gap) CVs, short-$P_{\rm orb}$ CVs, and period-bouncer CVs, respectively. 
Among our samples, although WZ Sge-type DNe have orbital periods and mass ratios near the period minimum, the majority of known WZ Sge-type DNe are \textit{not} a period bouncer (candidate). Indeed, \citet{pal20GaiaCVdensity} used the scale height of 260 pc for WZ Sge, V455 And, and GW Lib. Polars are typically found below the period gap \citep[e.g. ][]{sch25polar} and the orbital periods of our polar samples are below 2 hours. The orbital period distribution of systems showing micronovae is somewhat unknown, but given the orbital periods of the systems showing micronovae \citep[i.e. TV Col and EI UMa;][]{sca22micronova}, they most likely have orbital periods above the period gap.
Thus, we used \yt{the scale heights of} 260, 260, and 120 pc for WZ Sge stars, polars, and micronovae, respectively.}

\subsubsection{{Uniform distribution}}
\label{sec231:uniform}


\yt{It also has been proved that CVs not belonging to the thin-disk population do exist; halo \citep[e.g.][]{pat08j1507, lee19ksp1611a}, stellar clusters \citep[][and references therein]{pie08dngc, mod20dn47tuc}, and nearby galaxies including the Magellanic clouds \citep[e.g.][]{sha03lmcdne}. Given the absolute magnitude of WZ Sge-type DNe and micronovae at maxima ($M_\text{r}\leq$4 mag), these subtypes of CVs can be detected up to $\simeq100$ kpc with LSST. 
Hence,  LSST will for the first time provide an opportunity for systematic studies of CVs of more common subtypes in these populations, in addition to classical novae \citep[e.g.][]{sha23m87novae}.
Since integrating these populations as an individual component into a mock observation is beyond the scope of our paper, we consider a uniform and random distribution on the celestial sphere, including regions outside of the LSST footprints (the uniform distribution model). We also set a uniform and random distribution on the distance between 0.1 -- 100 kpc.} \footnote{We choose a uniform distribution on distance $d$ ($\propto d$) rather than the square of distance ($\propto d^2$; i.e., uniform density) to preserve enough numbers of nearby ($\leq$5 kpc) systems for more accurate measurements of detection efficiency across the wide range of distance.}
\yt{This model gives a uniform and statistical observation efficiency across the entire telescope pointings and distances beyond the thin disk population.}

\section{Results}
\label{sec:result}

\subsection{Overall performance}
\label{sec31:overall}

\begin{table}[tb]
    \centering
    \caption{Summary of WZ Sge star simulations using OpSim v4.3.1. }
    \begin{tabular}{cc}
    \hline
    Simulated & 20000\\
    \hline
    In LSST footprint & 19419\\
    Detection & 4014\\
    Detection in outburst & 3917\\
    Detection in the first 7 days in outburst & 2898\\
    With $r$-band quiescence counterpart & 43\\
    \hline
    \end{tabular}
    \label{tab:yt-ugwz_sim}
\end{table}

\begin{table}[tb]
    \centering
    \caption{Summary of polar simulations using OpSim v4.3.1. }
    \begin{tabular}{cc}
    \hline
    Simulated & 20000 \\
    \hline
    In LSST footprint & 19456\\
    Detection  & 1394\\
    $\geq$ 10 detections  & 918\\
    $\geq$ 100 detections  & 266\\
    $\geq$ 100 detections in $r$ band  & 100\\
    \hline
    \end{tabular}
    \label{tab:yt-amher_sim}
\end{table}

\begin{table}[tb]
    \centering
    \caption{Summary of micronova simulations using OpSim v4.3.1. }
    \begin{tabular}{cc}
    \hline
    Simulated & 20000 \\
    \hline
    In LSST footprint & 19450 \\
    Detection in quiescence  & 5509\\
    Detection in a micronova  & 144\\
    \hline
    \end{tabular}
    \label{tab:yt-mne_sim}
\end{table}


\begin{figure*}[t]
    \centering
    \includegraphics[width=\textwidth]{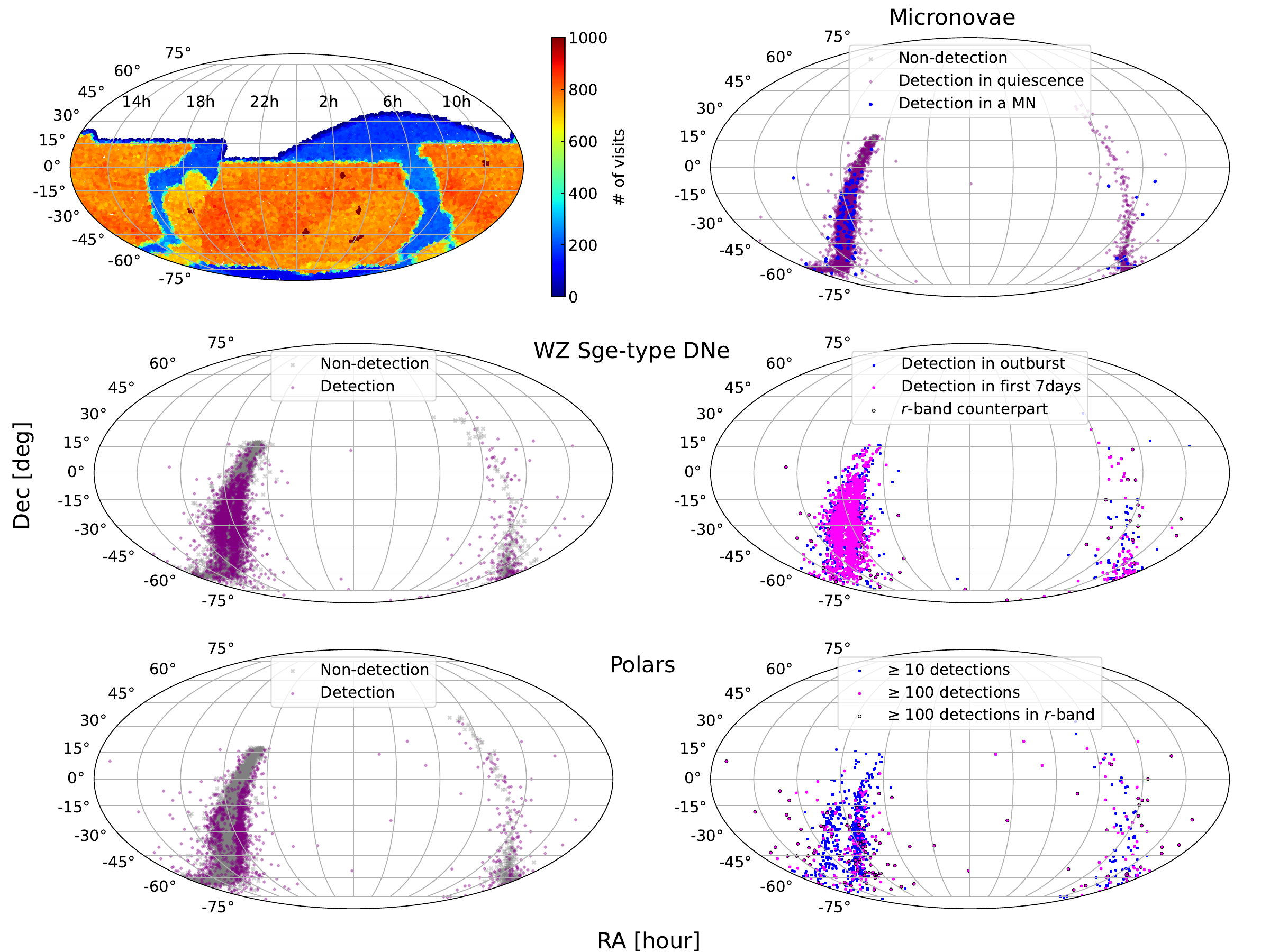}
    \caption{
    Top left; The number of the LSST visits in the whole 10 years.
    Detectability dependence on the equatorial coordinate for Micronovae (upper right), WZ Sge stars (middle) and AM Her stars (lower) \yt{of the thin-disk distribution model}. The description of each plot and category is presented in text.}
    \label{fig:yt-skydist-TD}
\end{figure*}


\begin{figure*}[t]
    \centering
    \includegraphics[width=\textwidth]{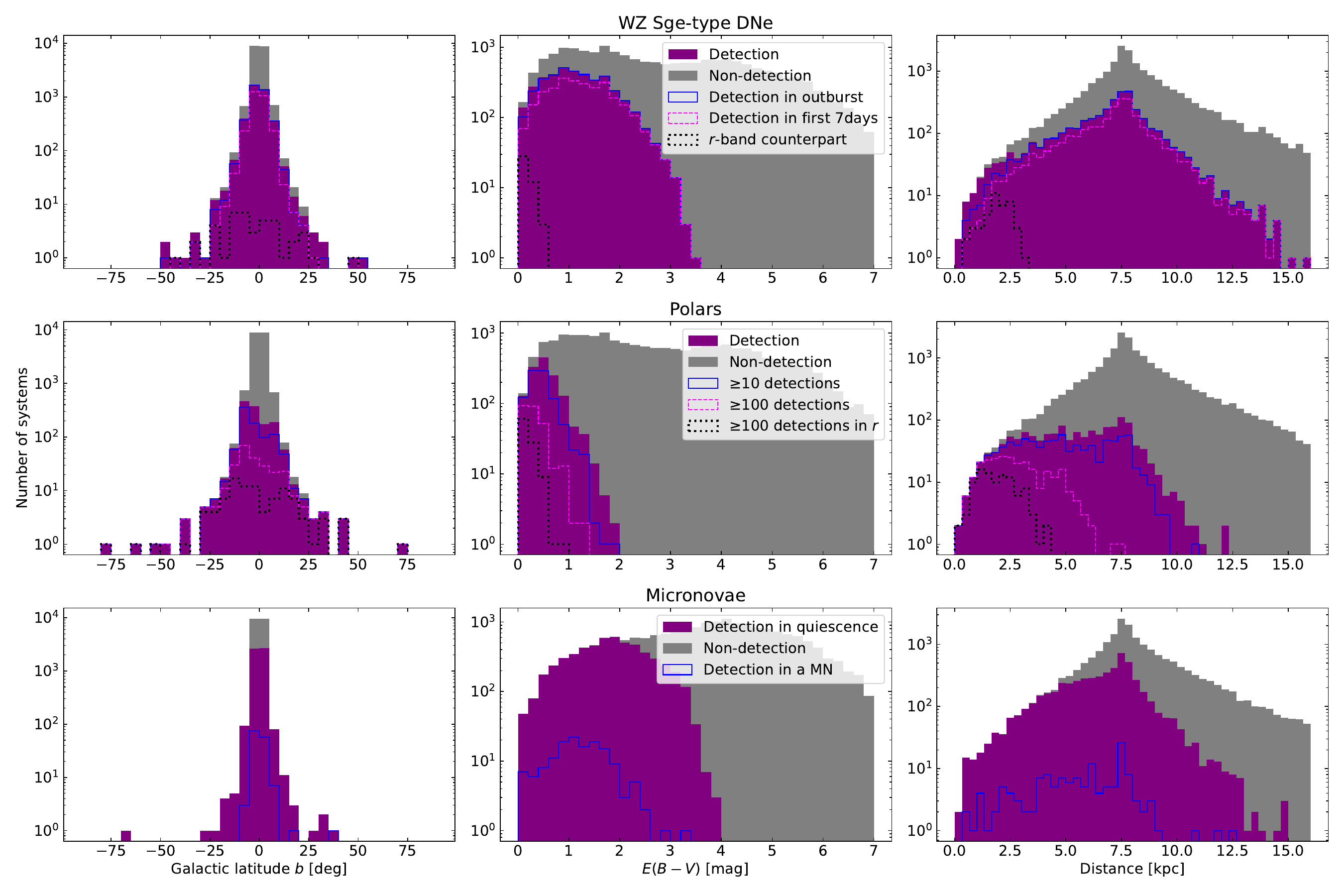}
    \caption{Numbers of detection categories on various system parameters \yt{in the thin-disk distribution model}; Galactic latitude $b$ (deg) (left column), Galactic extinction $E(B-V)$ (mag) (middle column), and distance (kpc) (right column) of WZ Sge-type DNe (top row), polars (middle row), and micronovae (bottom row).}    \label{fig:yt-detectability-TD}
\end{figure*}

Table \ref{tab:yt-ugwz_sim},  \ref{tab:yt-amher_sim}, and  \ref{tab:yt-mne_sim} summarizes the results of our simulations \yt{with the thin-disk distribution model}. The examples of our simulated light curves are presented in Appendix Figure \ref{fig:yt-samplelcs}.
\yt{Among 20000 simulated samples of each subclass, 19450 are in the footprint of the LSST.}
For the WZ Sge stars, we further categorize the samples based on their detections in and outside of an outburst; a sample has no detection neither in outburst nor quiescence, at least one detection either in outburst and/or quiescence, detection in outburst, detection within the first seven days in outburst, and detection both before the outburst in $r$ band and within the first seven days in outburst. 
The last samples represent those that will have a quiescence counterpart in individual visits, hence the outburst amplitude can be estimated. 
The polars are further categorized based on the number of detections; a sample has no detection, at least one detection, $\geq 10$ detections, $\geq 100$ detections, and $\geq 100$ detections solely in $r$ band over 10 years. The $\geq 100$ detections is a rather arbitrary value. Given the CRTS coverage of 200--400 measurements over $\geq$ 7 years and 4 measures per day (within a $\sim$ 30 min window) every $\sim$2 weeks, detections of an order of 100 would be enough to identify and study the state changes of polars \citep{khinethesis}.
For micronovae, 
{we first choose systems with at least one detection in quiescence.}
We set our criteria for the number of detections during a micronova eruption as any detections brighter than 3$\sigma$ deviation from the quiescence level. This still gives the most optimistic way since (i) we assumed a constant brightness in quiescence and hence the deviations are determined by the telescope performance, and (ii) 
\yt{CVs in general}
may show intrinsic orbital modulation and some other types of outbursts and state transitions which will enlarge the deviation of its quiescence brightness.  
\yt{We categorized our samples into; a sample with at least one detection in quiescence and with at least one detection during a micronova in addition.}

The top left panel {of Figure \ref{fig:yt-skydist-TD}} presents the total number of visits in the 10-year LSST operation. 
\yt{The top right and lower panels of Figure \ref{fig:yt-skydist-TD} present the spatial dependence of our source categories on the celestial sphere in equatorial coordinates from the thin-disk distribution model.
}
Figure \ref{fig:yt-detectability-TD} represents the numbers of the samples in our detection categories of WZ Sge stars, polars, and micronovae, from top to bottom, respectively, on the Galactic latitude $b$ (deg) (left), Galactic extinction $E(B-V)$ (mag) (middle), and distance (kpc) (right). 
Samples in the thin-disk distribution accumulate towards the Galactic center at Galactic latitude $|b| \leq 5$ deg, distance around 8 kpc, and Galactic extinction $E (B-V) \geq 1.0$. 
One clearly sees less detection along the Galactic plane due to higher Galactic extinction and fewer visits, except the Galactic center areas included in the WDF. 

\yt{Given their bright outburst maximum, WZ Sge stars in outburst can be observed even beyond 8 kpc towards the less extincted directions.}
On the other hand, considering the distance modulus of $\approx 13.5$ mag at 5 kpc and {$r$-band absolute magnitude} $M_{r} \sim 11.5$-mag in quiescence, their $r$-band counterpart is only detectable up to 5 kpc in the individual images \yt{without considering the extinction.}
\yt{Along the Galactic plane, this is reduced due to higher Galactic extinction.} 
\yt{Even if an outburst is missed due to seasonal gaps or its poor cadence, LSST would be capable of detecting more than 50\% of such intrinsically faint CVs around the period minimum in quiescence at 2.0 kpc. These can be identified as a CV via their orbital and other variability, as proved with ZTF \citep[][]{van22ztfamcvn, gal24j0411}, although this is beyond the scope of this paper.}
We do not expect any WZ Sge-type DNe even in outburst {in the directions and distances for {$E(B-V) \geq 3.5$}}. 
Polars are fainter at their high state compared to WZ Sge-type DNe, but their high states usually last longer. 
\yt{The fraction of polar samples with $\geq 100$ detections drops below 50\% beyond 2 kpc.}
{The high extinction towards the Galactic center results in almost no detection of polars beyond 10 kpc}.
Compared to these, micronovae are intrinsically brighter and \yt{show a less dependence on distance. Almost no micronovae are expected beyond 10 kpc due to the decreased number of detected systems in quiescence.}
The short duration of the eruption, however, results in a much less efficient detectability; 
2.6\% of samples with at least one detection in quiescence are detected during the eruption. 
\yt{
It must be noted that our micronova criterion requires a quiescence counterpart, and hence heavily depends on the absolute quiescence brightness of the system.
With the same burst luminosity, we expect a higher (lower) detection fraction for a system with intrinsically fainter (brighter) quiescence in absolute magnitudes because of a longer (shorter) effective burst duration and a larger (smaller) burst amplitude.}

\subsection{WZ Sge stars in outburst}
\label{sec32:wzsge}

\begin{figure}[t]
    \centering
    \includegraphics[width=\columnwidth]{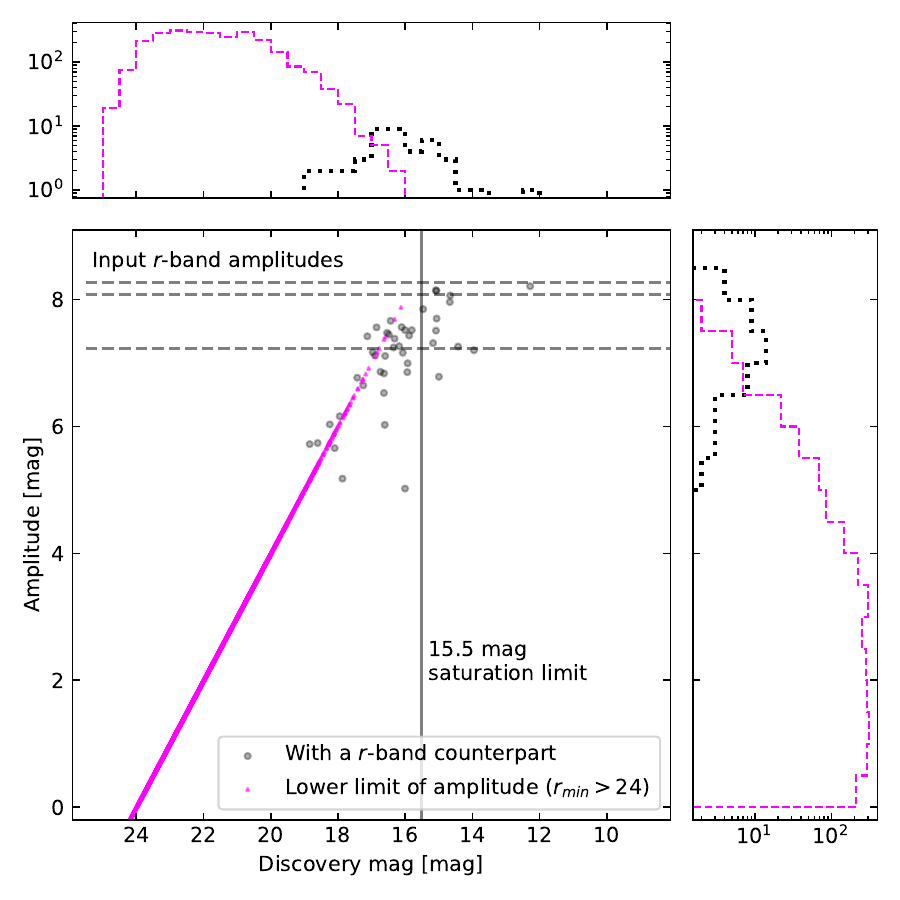}
    \caption{Observed magnitude at the first detection in outburst vs the outburst amplitude from the $r$-band counterpart in the simulated light curves of WZ Sge-type DNe. {The vertical and horizontal lines show the typical saturation magnitude in the LSST and the outburst amplitudes of the input light curves, respectively.}
    In case there is no detection before the outburst in the $r$ band, we assume the upper limit $r_{\rm low} \geq 24.0$ mag and give a lower limit on the outburst amplitude. Only the samples  {in the thin-disk distribution model} whose outbursts are detected within the first 7 days are plotted.}
    \label{fig:yt-wzsge}
\end{figure}

\begin{figure*}[t]
    \centering
    \includegraphics[width=\textwidth]{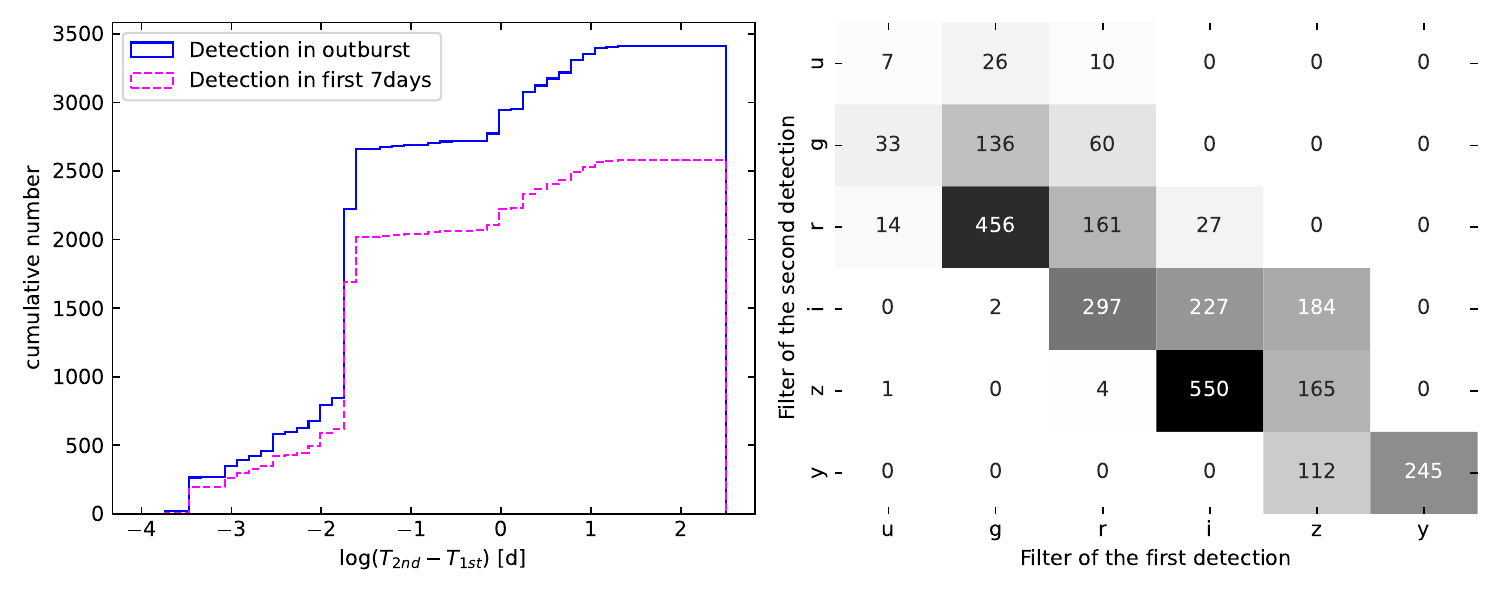}
    \caption{Left; Cumulative histogram of the time difference between the epochs of the first and second detection in outburst of the simulated WZ Sge-type DNe {in the thin-disk distribution model}. Only samples with detections in outburst are used for the plot.
    Right; the number of samples with the filters of the first (x-axis) and second (y-axis) detections in outburst. Samples with two or more detections on the outburst discovery night are used in plot.}
    \label{fig:yt-wzsge-peak}
\end{figure*}

Among the simulated WZ Sge-type DNe 
\yt{in the thin-disk distribution model}, 
3917 (20\%) objects have at least one detection during an outburst.
2898 (15\%) objects have a detection within the first seven days in outburst. Among these first-seven-day samples, the $r$-band quiescence counterpart in individual scans exists only in 
43 (1.4\%) samples.  
\yt{Thus, the majority of their outbursts will have only a lower limit on the amplitude.} We note that, in later years of the LSST operation, more systems would have a counterpart in the deeper co-add images in future data releases.


Figure \ref{fig:yt-wzsge} represents the magnitude at the first detection after the beginning of an outburst regardless of the band vs the amplitude of the outburst based on the median $r$-band magnitudes before the outburst {in the thin-disk distribution model}. In the cases where there is no pre-outburst counterpart, we set the lower limit on the outburst amplitude assuming the quiescence counterpart is fainter than 24.0 mag, based on the typical limiting magnitude in the $r$ band with a single visit. 
The typical saturation limit of the LSST is $\approx 15.5$ mag in all bands \citep{lsstsciencebookv2}. Although the nearest \text{(within 2 kpc with the distance modules of $\leq$11.5 mag)} WZ Sge-type DNe in outburst can be saturated in LSST, such bright outbursts should be discovered by other shallower and faster transient surveys. 
This figure clearly illustrates that outbursts 
of discovery magnitude $\leq 17.5$ mag will have information on the amplitude of the outburst and probably the color of the quiescence counterpart. This is expected given the limiting magnitude of 24--25 mag in the LSST, a typical outburst amplitude of 7--9 mag, \yt{and a rise timescale of $\leq0.2$ d mag$^{-1}$} in WZ Sge-type DNe. Therefore, WZ Sge-type DNe fainter than 17.5 mag at discovery \yt{are indistinguishable from both Galactic and extragalactic fast transients solely based on the lower limit of the outburst amplitude.}
Other information such as rise/decline timescales, outburst duration, outburst cycle, color, and existence of a possible host galaxy is required to classify them based on the photometric data, otherwise an optical spectrum is needed for classification.

In order to conduct proper follow-up observations and classifications, the observation results around the outburst maximum are vital.
The left panel of Figure \ref{fig:yt-wzsge-peak} presents the time delay between the epochs of the first $T_{\rm 1st}$ and second $T_{\rm 2nd}$ detections in an outburst.
Out of 3917 systems detected in outburst, 2717 (69\%) systems will have their second detection within the same night {and specifically within 45 min for 2657 samples}. 1776 (65\%) of them will be observed in different bands in their first and second detection in outburst, while the remaining 941 will be observed in the same band (right panel of Figure \ref{fig:yt-wzsge-peak}). Hence, if a one-hour delay of follow-up observations is allowed, which is presumably not problematic for transients with days--weeks timescales like dwarf nova outbursts and for observatories at different longitudes, we expect that multiple-detection alerts will dominate, not only enabling the evaluation of either a color or a rise/decline rate, but also rejecting bogus events.

\subsection{Polars}
\label{sec33:polar}

\begin{figure*}[t]
    \centering
    \includegraphics[width=\textwidth]{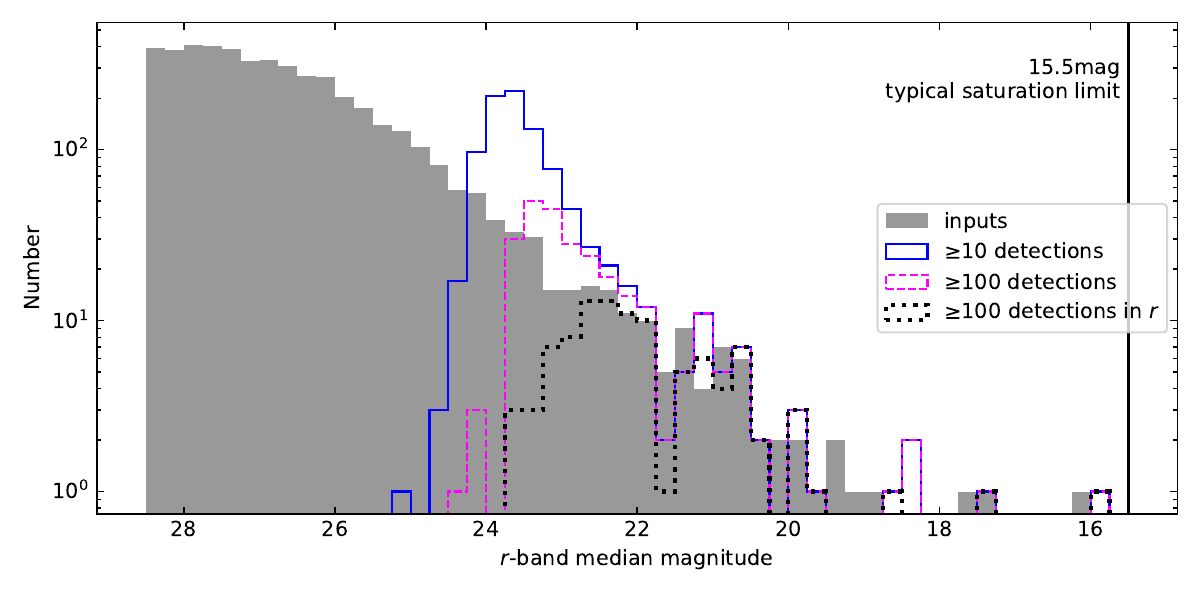}
    \caption{Histogram of the $r$-band median magnitude of our polar samples in the 10-years simulation {under the thin-disk distribution model}. Gray histogram represents that of the input light curves, while blue, magenta, and black histograms represent those of the simulated samples with $\geq$ 10, $\geq$ 100, and $\geq$ 100 in $r$ band detections.}
    \label{fig:yt-polars}
\end{figure*}


\yt{Among the simulated samples of polars in the thin-disk distribution model, }
{1394, 918, and 266 (7.2, 4.7, 1.4\%) samples have $\geq$ 1, 10, and 100 detections, respectively.} \yt{These lower numbers than WZ Sge-type DNe are resulted} from the high concentration of the simulated samples at the Galactic center and their intrinsic faintness.
Figure \ref{fig:yt-polars} represents the $r$-band median magnitude distribution of the input (gray) and simulated (lines) light curves {in the thin-disk distribution model}. It is clear that systems brighter than 22.5 mag in median (thus varying in 24--20 mag) have a similar distribution between the input and simulated, meaning that the multiple states in polars are well covered. On the other hand, the median magnitude of systems fainter than 22.5 mag will be biased towards the brighter systems; their faint state is not detected and only a partial light curve in high state will be obtained.

\subsection{Micronovae}
\label{sec34:micronova}

\begin{figure*}[t]
    \centering
    \includegraphics[width=\textwidth]{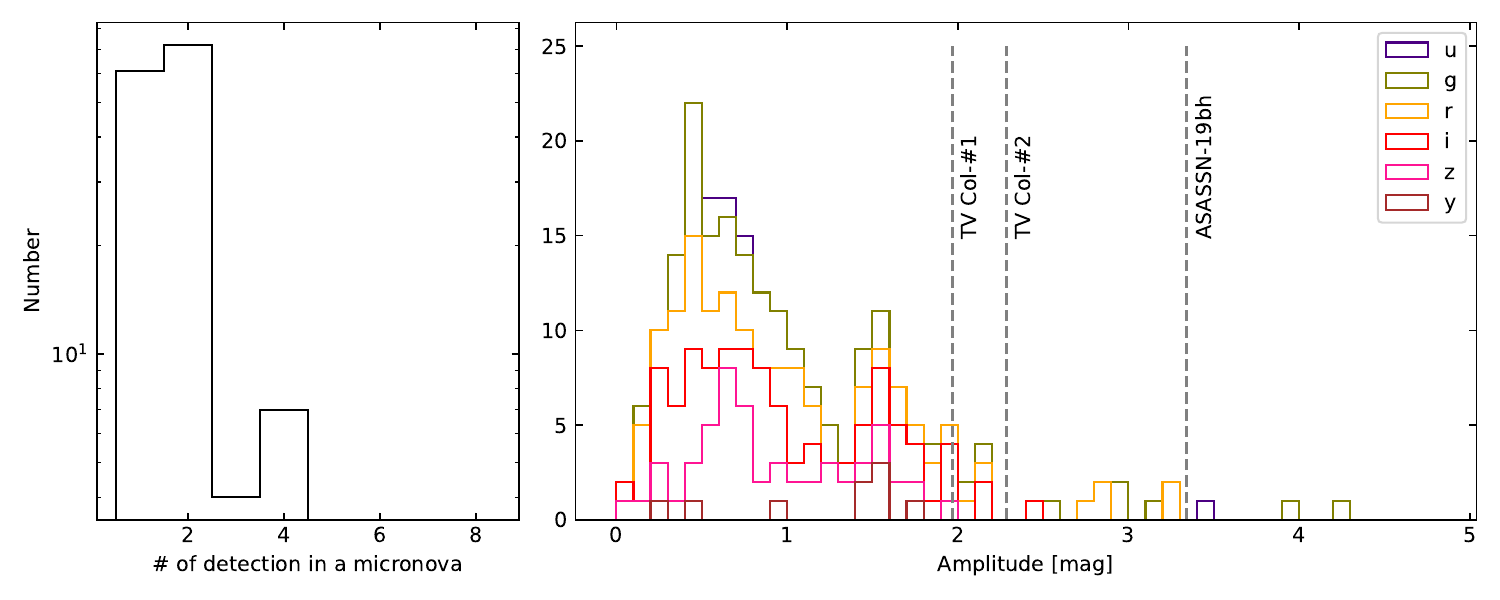}
    \caption{Left; histogram of number of detections during a micronova regardless of the filters {in the thin-disk distribution model}. Right; histogram of the simulated amplitudes of micronovae. The vertical lines represent the $r$-band amplitude of the three input micronovae.}
    \label{fig:yt-micronovae}
\end{figure*}

Micronovae are a short-lived event within 1 day, much shorter than DNe and polars. This results in a smaller fraction of micronova eruptions detected; 
144 (2.6\%) samples.
The left panel of Figure \ref{fig:yt-micronovae} represents the histogram of the number of detections during the micronovae event {in the thin-disk distribution model}. 
Because there will be two or more visits per footprint on the same night in the WDF survey, {more than half} of micronovae will have multiple detections, and be observed in different filters. Since current \yt{detailed light curves} of micronovae are limited in \yt{the TESS observations} with single and broad band, this planned cadence strategy can benefit the study of the color of micronovae and origin of these fast eruptions.
The right panel of Figure \ref{fig:yt-micronovae} presents the distribution of the simulated amplitude of micronovae {in the thin-disk distribution model}. The amplitude distribution peaks below 1.0 mag, compared to 2--4 mag amplitude in the input light curves. Thus it is clear that the observed amplitudes are underestimated by 1--2 magnitudes in most cases because of their faster timescale and poorer cadence in the LSST compared to those in TESS. 
\yt{Since our detection criteria are rather simplified, the actual recovery rate is likely to be lower because of factors such as state changes and orbital modulation with an amplitude of $\geq 0.1$ mag generally observed in CVs.}
A more sophisticated detection method may be needed to efficiently detect such fast events in CVs.

\subsection{\yt{Implications from the uniform distribution model}}
\label{sec35:uniform_model}

\begin{figure*}[t]
    \centering
    \includegraphics[width=\textwidth]{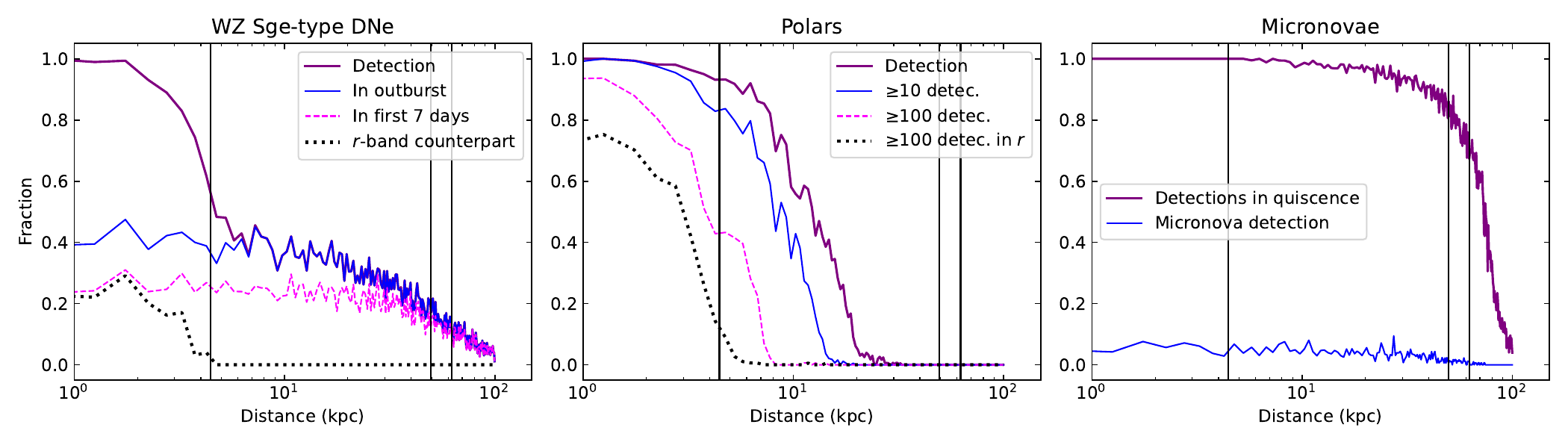}
    \caption{\yt{Detection efficiency dependence on distance according to the uniform distribution model, for WZ Sge-type DNe, polars, and micronovae from left to right. The colors have the same meaning as Figure \ref{fig:yt-skydist-TD}. The vertical lines indicate the distance of 47 Tuc, LMC, and SMC.}}  
    \label{fig:yt-detectability-Uni}
\end{figure*}

\yt{
As mentioned in Section \ref{sec231:uniform}, LSST is capable of studying many CV subtypes beyond the thin disk populations. Based on our uniform distribution model, Figure \ref{fig:yt-detectability-Uni} presents the normalized fraction of respective detection categories (WZ Sge-type DNe, polars, and micronovae from left to right) on distance. The vertical lines indicate the distances of 47 Tuc \citep[4.45 kpc;][]{che1847tucngc362_dist}, the Large Magellanic Cloud \citep[49.6 kpc;][]{pie19lmc_dist}, and the Small Magellanic Cloud \citep[62.4 kpc;][]{gra20smc_dist}. More than 90\% of the simulated samples are located at the direction and distance with $E(B-V) \leq 0.5$ mag, unlike the thin-disk distribution model.
Among the WZ Sge-type DN samples, all samples are detected at least once up to 1.5 kpc. The fraction of outburst detections in this range is $\simeq$40\%, which is primarily determined by the LSST cadence. The fraction of outbursts with an $r$-band counterpart decreases beyond 1.5 kpc, and we do not expect any counterpart beyond 4.5 kpc where a distance modulus is $\simeq13.2$ mag. 
The 10-year full data release, which is expected to reach $\simeq27.5$ mag \citep[][]{lsstsciencebookv2}, can push this limit to 10 kpc, including 47 Tuc.
Beyond these distances, LSST is capable of detecting WZ Sge-type DNe only in outburst. For example, the detected fractions are 35\%, 17\%, and 14\% at the distances of 47 Tuc, LMC, and SMC, respectively. The absolute magnitudes of transients from these populations are easily estimated, helping the classification e.g. between dwarf novae and classical novae.
Particularly for globular clusters, theoretical models predict a higher fraction of short-orbital-period CVs than local volume-limited samples because of their old and burst-like star formation history \citep[e.g.][]{bel16gccvs}. The contentious baseline of LSST for 10 years can help to identify such an evolved DN undergoing an outburst.
%
The longer but fainter high state of polars results that only within 2 kpc $\simeq$90\% of the samples have $\geq100$ detections. The fraction of the $\geq100$-detection samples at the distance of 47 Tuc is 36\%. The fraction of the $\geq10$-detection samples becomes below 50\% beyond 9.0 kpc. Hence, even with LSST, polar populations beyond the thin-disk ones will not be well constrained.
The quiescence of our micronova samples will be successfully detected up to 35 kpc with $\geq90$\% rate. The recovery rate ($\simeq$4\%) of micronovae does not show a significant change up to this distance, because of their even brighter burst than in quiescence. 
Its worth noting that, if a micronova luckily occurs within the Deep Drilling Fields or the ToO Fields, although an expected number of CVs at such high Galactic latitude should be small, the eruption can be observed up to $\approx90$ visits but within one night (typically within $\sim$ 70 min in different bands; right-bottom panel in Figure Appendix \ref{fig:yt-samplelcs} for example). 
}

\section{Discussion}
\label{sec:discussion}

\subsection{LSST capability for CV studies}
\label{sec51:lsstcv}


As we have explored in the previous section, LSST will be capable of detecting many subtypes of cataclysmic variables at an enormous \yt{fraction as seen above} thanks to its very deep limiting magnitude and consequently larger volume to be observed.
Our simulation for WZ Sge stars shows that about {20\%} of their outbursts will be detected by the LSST {under the realistic thin-disk distribution in our Galaxy}. 
Since their typical outburst cycles are the order of a decade, the 10-year operation of the LSST is long enough to study a certain fraction of WZ Sge stars within the footprint of the LSST, even beyond 1 kpc where a counterpart has not been detected to date.
\yt{In the thin-disk distribution model, 36 samples have the simulated outburst maximum $r\leq$15.0 mag and declination higher than $-30^\circ$, detectable with ZTF. According to the Variable Star Index \citep[VSX;][]{VSX}, there are 57 WZ Sge-type DNe and candidates whose outburst maximum is brighter than 15.0 mag and which are registered with an associated ZTF alert name over its 7-year operations. This number must be considered as a lower limit because (A) there are some famous systems which has the assigned ZTF alert name but it is not registered in VSX (e.g., OV Boo, RZ Leo, AY Lac, KK Cnc, and V748 Hya) and (B) some well-confirmed bright WZ Sge-type DNe do not even have any associated ZTF alerts \citep[e.g. TCP J23580961+5502508 (vsnet-alert 26954\footnote{\url{http://ooruri.kusastro.kyoto-u.ac.jp/mailarchive/vsnet-alert/26954}}); TCP J16271026-1030020 (vsnet-alert 28094\footnote{\url{http://ooruri.kusastro.kyoto-u.ac.jp/mailarchive/vsnet-alert/28094}}); TCP J09370380+1657350;][]{iso21j0937_atel}}.
\yt{This yields that at least $\simeq 9$ WZ Sge-type DN outbursts brighter than 15.0 mag have been in the observed fields of ZTF every year.
Thus, by interpolating these numbers; multiplying by a factor of 0.25 in Table \ref{tab:yt-ugwz_sim}, the planned LSST cadence is capable of reporting at least a few new WZ Sge stars in outburst every night and more than one thousand per year.
}
The limiting magnitude of ongoing transient surveys is $\approx$ 20 mag. Thus it will still be challenging, especially in the initial years of the LSST, to identify the subtype of DNe peaking fainter than 20 mag and lacking a prior counterpart \yt{solely from the LSST light curve}, because there will be no information on outburst cycles.


LSST will provide \yt{light curves with $\geq$100 detections of polars within 2 kpc and sparser ones with $\geq$10 detections at $\leq$8 kpc along the Galactic plane}. 
{Given that 80 systems are listed as polar within 500 pc in \citet{sch25polar}, which is only 7 samples in our polar simulations, LSST will provide the light curves of an order of thousands of Polars with more than 100 detections over the 10 years.}
Fast events, like micronovae, have a lower detection fraction due to the relatively long cadence of the LSST. However, if such a fast event occurs within a Deep Drilling Field, time series observations can be obtained with the LSST, probing their multi-color nature. 

{Our simulations still assume distributions following the simple Galactic thin disk (thin-disk distribution model). Simulations with the intrinsic space density, system fraction between subtypes, realistic recurrence time of outbursts and, more detailed Galactic structures (e.g. the Galactic bulge, globular clusters, and Magellanic clouds) are beyond the scope of this paper and we leave to future works.
}


\subsection{Application to other CVs and transients}
\label{sec52:others}

\begin{figure}[t]
    \centering
    \includegraphics[width=\columnwidth]{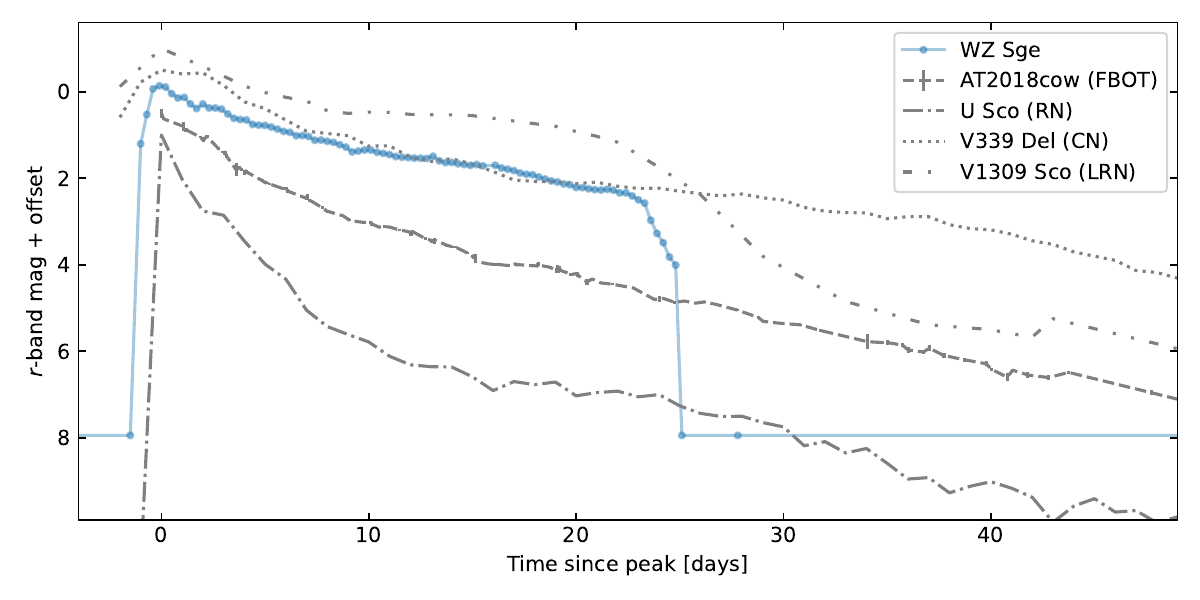}
    \caption{Light curve comparison of WZ Sge \citep{kat09pdot}, recurrent nova U Sco \citep{mur24usco}, classical nova V339 Del (from AAVSO LCGv2\footnote{\url{https://www.aavso.org/LCGv2/}}), FBOT AT 2018cow \citep{pre18cow}, and LRN V1309 Sco (from AAVSO LCGv2).}
    \label{fig:yt-wzsgecomp}
\end{figure}

We have explored the various discovery spaces of cataclysmic variables at various timescales. Thus our simulations are not only applicable to the limited subtypes of CVs but can be used to characterize and create an observation strategy on various types of Galactic and extragalactic transients. 
Figure \ref{fig:yt-wzsgecomp} compares the light curves of {Galactic and extragalactic fast transients comparable to WZ Sge}.
WZ Sge stars typically rise within 1--2 days and decline with a timescale of one week per mag. These are comparable to those of AM CVn stars in outburst \citep[e.g.;][]{roe21ztfamcvn}, which are explained by the same disk instability model as hydrogen-rich DNe but their disk is helium-rich \citep[e.g.;][]{tsu97DIamcvn}.
Since the light curve shape of WZ Sge-type DNe and long-period ($\geq 30$ min) AM CVn stars in outburst are roughly identical, especially around the outburst maximum, they will share a similar discovery space but the expected number of outbursts is orders of magnitude lower in AM CVn stars based on the current transient surveys (see e.g. VSX).
Classical novae also show similar large amplitude, fast rise, and week-timescale decline \citep{kaw21DNCNinASASSN}. The expected Galactic nova rate is however much lower; an order of $\approx50$ per year \citep[e.g.][]{de21PGIRnova}. 
Some fast extragalactic transients, such as fast blue optical transients \citep[FBOTs; e.g.,][]{pre18cow} and luminous red novae \citep[LRNe; e.g.,][]{pas19lrnreview}, also show a similar fast rise and decline, while they should be accompanied by a detectable host galaxy in LSST images and appear independent of the Galactic latitude.

As we see in section \ref{sec33:polar}, polars brighter than $\approx$ 22.5 mag on average will be covered well enough for e.g. multi-state studies, while fainter systems are only observed in their higher states. The typical range of brightness changes in our polar samples are 2--4 mag. This is similar to those of normal DN outbursts and VY Scl-type dimming in novalike variables \citep[][]{Warner1995}. Thus, the unbiased long-term trends of the majority of CVs also can be studied up to $\approx 22$ mag.

Although the expected recovery fraction of micronovae is low ($\approx 2.6 \%$), the deep and large field of view of the LSST will be capable of detecting a reasonable number of such day-scale events in CVs and other transients. These include stellar flares from the secondary stars of CVs \citep{sch22v2487oph} and from various types of stars \citep[e.g.,][]{mae12stellarflare}.
The important lesson from our micronovae cases is that the observed amplitude (or peak brightness) can always be underestimated by 1--2 magnitudes because of the cadence of the LSST. Meanwhile, if such a fast transient {and a fast rise of dwarf nova outbursts} take place during the Deep Drilling Field surveys, more than 5 scans will be obtained within one night, enabling the detailed multi-color analysis. The fast publication of transient alerts in DDFs can even enable coordinating fast response and simultaneous follow-up observations with other telescopes.

\section{{Summary}}
\label{sec:summary}
{We performed systematic mock observations of CVs with LSST, to investigate its capability for detecting CVs with various timescales and brightness. 
We simulated the multi-color optical light curves of WZ Sge-type DNe, multi-state polars, and micronovae \yt{(so far seen in a few IPs only)] under the thin-disk approximation of the distribution of CVs in our galaxy. Our key findings are summarized as follows;}

\begin{itemize}
    \item 
    {Only 20$\%$ of WZ Sge-type DNe \yt{in the thin-disk population} will be detected during outbursts by LSST because of higher extinction and concentration along the Galactic plane.  Only those discovered at brighter than 17.5 mag at outburst maximum will have a \yt{quiescence} counterpart in previous LSST epochs. \yt{WZ Sge-type DNe in quiescence  at 2 kpc can be detected by LSST at a chance of $\simeq$50\%, even if the outburst was missed.} 
    70$\%$ of WZ Sge-type DN outbursts will have their second detection on the discovery night with the LSST cadences currently planned, and two-thirds will be observed in different bands. Thus, color information on the discovery night could be useful to distinguish transients with similar light curve properties, including AM CVn stars, classical novae, and fast extragalactic transients.}

    \item 
    \yt{More than half of polars up to 2 kpc can be covered with $\geq 100$ detections along the Galactic plane with the 10-year LSST WFD survey.} LSST will provide the light curves covering multiple states for such systems, with a median brightness of 22.5 mag or brighter; only the bright state will be covered by LSST for fainter systems. Normal DNe and novalike stars brighter than 22 mag, which have a similar range of variability, will also get good coverage to check their long-term variability.

    \item 
    {The 1-d timescales of micronova eruptions result in a lower recovering rate of 2.6$\%$, and their amplitudes will be underestimated by $\simeq 1$ mag. The multiple visits within the same night with different filters may reveal the color and SED information of micronovae, which is currently unknown. Such a fast eruption happening during DDF visits potentially provides information on their color evolution, although the chances are very low.}

    \item 
    \yt{Our uniform distribution model indicates that intrinsically bright CVs and high states like DN outbursts and micronova bursts can be detected even at the distances of globular clusters and Magellanic clouds, providing an important opportunity to prove their CV populations. At the distances of 47 Tuc and LMC, 35 and 17\% of WZ Sge-type DN outbursts can be detected by LSST. On the other hand, the intrinsic faintness of polars results in the $\geq10$-detection samples being limited within 10 kpc. The known distances of globular clusters and Magellanic clouds may also help to identify the CV subtypes based on the absolute magnitudes.}

\end{itemize}

\section*{Acknowledgements}

We thank Rachel Street and others in the Rubin LSST Transient and Variable Star collaboration for useful inputs and advice.

DAHB acknowledges research support from the National Research Foundation, which supports LSST participation for JPM, DAHB, SM, YT and AvD. ML acknowledges support from the South African Radio Astronomy Observatory and the National Research Foundation (NRF) towards this research. Opinions expressed and conclusions arrived at, are those of the authors and are not necessarily to be attributed to the NRF.



\bibliography{bibs}{}
\bibliographystyle{aasjournal}




\appendix

\begin{figure*}[h]
    \centering
    \includegraphics[width=\textwidth]{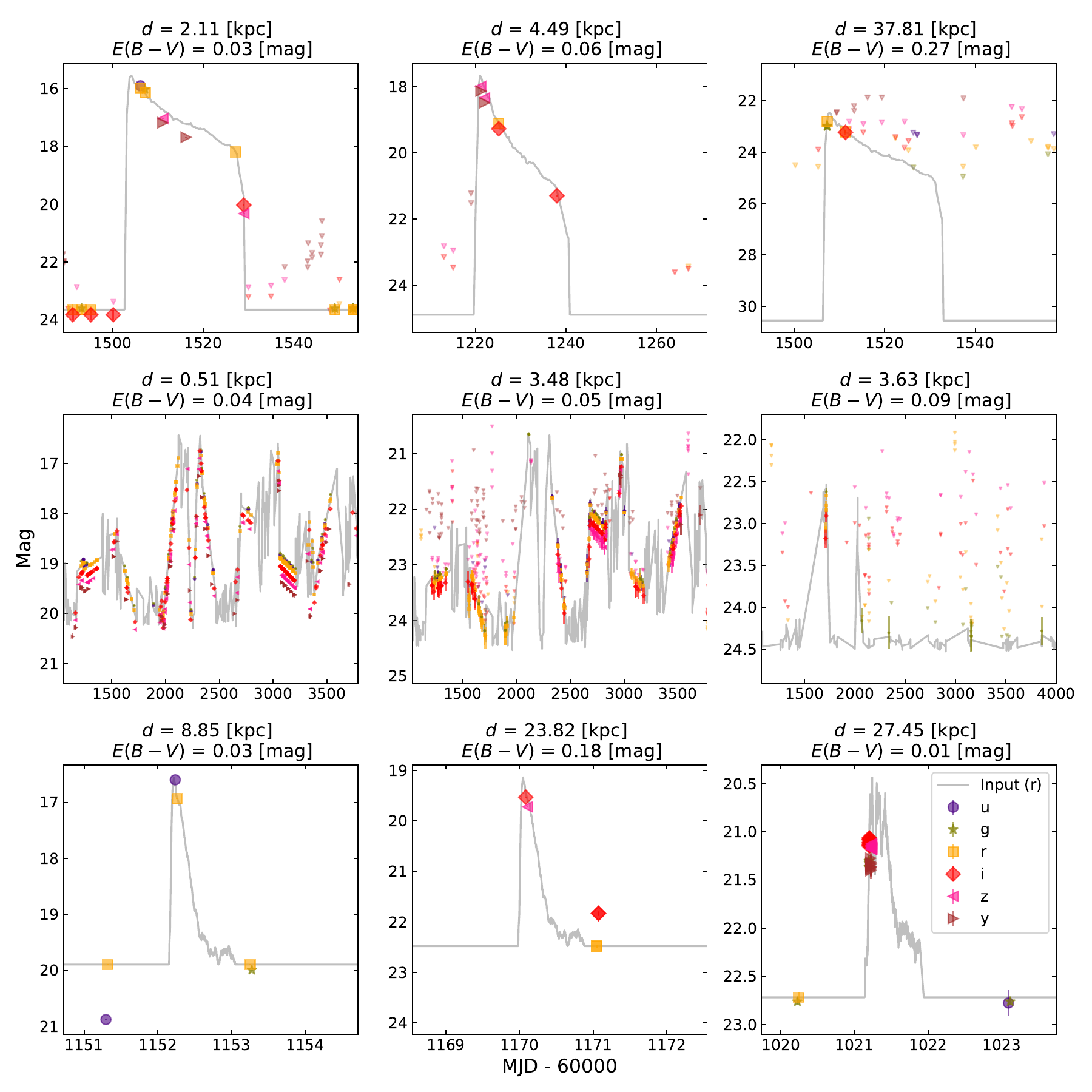}
    \caption{Examples of our simulated light curves of WZ Sge stars, polars, and micronovae from top to bottom. The gray line represents the input light curve in $r$ band, only adjusted to the distance modulus. The inverse triangle means the upper limits.}
    \label{fig:yt-samplelcs}
\end{figure*}


\end{document}